\documentclass[10pt,a4paper,twocolumn]{article}

\begin{document}

\title{To string together six theorems of physics by Pythagoras theorem}
\author{H. Y. Cui\\
Department of Applied Physics\\
Beijing University of Aeronautics and Astronautics\\
Beijing, 100083, China}

\date{( May, 8, 2002 )}

\maketitle

\begin{abstract}
In this paper, we point out that there are at lest six theorems in
physics sharing common virtue of Pythagoras theorem, so that it is possible
to string these theorems together with the Pythagoras theorem for physics
teaching, the six theorems are Newton's three laws of motion, universal
gravitational force, Coulomb's law, and the formula of relativistic dynamics
. Knowing the internal relationships between them, which have never been
clearly revealed by other author, will benefit the logic of physics teaching.
\end{abstract}

\section{Introduction}

If there is one mathematical theorem that is familiar to every university
student, it is surely the theorem of Pythagoras. The theorem was embedded in
physics like a gene even at the initiation of physics, because the physics
begins with describing the motion of a body in a frame of reference by
mathematical language---inevitably including the Pythagoras theorem.
Therefore, by seizing the Pythagoras theorem, we hope draw out some deeper
relationships in the physics.

Consider a particle of rest mass $m$ in our frame of reference $%
S(x_1,x_2,x_3,t)$, the particle moves a distance $\Delta l$ during
infinitesimal time interval $\Delta t$ with speed $v$, according to
Pythagoras theorem, we have

\begin{equation}
\Delta l_{}^2=v^2\Delta t^2=\Delta x_1^2+\Delta x_2^2+\Delta x_3^2
\label{p1}
\end{equation}
The above equation directly forms the velocity formula given by

\begin{equation}
v^2=v_1^2+v_2^2+v_3^2  \label{p2}
\end{equation}
Multiplying the above equation by the rest mass $m$, differentiating it with
respect to time, and defining a symbol $\mathbf{f}$ as

\begin{equation}
\mathbf{f}=\frac{d(m\mathbf{v})}{dt}  \label{p3}
\end{equation}
From Eq.(\ref{p2}), we obtain

\begin{equation}
\frac{d(\frac 12mv^2)}{dt}=\mathbf{f}\cdot \mathbf{v}  \label{p4}
\end{equation}
From the above two equations, we have seen what we want; when $\mathbf{f}$
represents the force exerting on the particle, Eq.(\ref{p3}) is the Newton's
second law of motion, and Eq.(\ref{p4}) is the kinetic energy theorem.

\section{An approach to Newton's three laws of motion}

We may rewrite the above section in an axiomatical way.

Axiom 1: Pythagoras theorem is valid only in inertial frame of reference.

From this Axiom we can derive out some useful consequences.

\textit{Consequence 1: The Newton's second law of motion can be derived from the
axiom 1, accompanied by deriving out the kinetic energy theorem for a
particle.}

\textit{Consequence 2: The Newton's first law of motion can also be derived from the
axiom 1, because of Eq.(\ref{p3}).}

Now we consider a composite system which contains two particles Bob and
Alice, whereas both Bob and Alice are also composed of many identical
constituent particles, the number of the constituents in Bob is $N$, the
number in Alice is $N^{\prime }$. The composite system of Alice and Bob can
be regraded as a single particle whose geometric center has a velocity $v_c$
given by

\begin{equation}
\mathbf{v}_c=\frac{N\mathbf{v}+N^{\prime }\mathbf{v}^{\prime }}{N+N^{\prime }%
}  \label{p5}
\end{equation}
Without the lost of generality, we have supposed that the $N$ constituents
of Bob have the same velocity $\mathbf{v}$, likewise for Alice with $\mathbf{%
v}^{\prime }$. Since the composite system of Bob and Alice can be regraded
as a single particle, it obeys the Newton's second law of motion as

\begin{equation}
\frac{d[(N+N^{\prime })m\mathbf{v}_c]}{dt}=\mathbf{f}_{ext.}  \label{p6}
\end{equation}
where $m$ is the mass of one constituent, $\mathbf{f}_{ext.}$ represents the
external force exerting on the composite system. If the external force
vanishes, then the system becomes

\begin{equation}
\frac{d(Nm\mathbf{v})}{dt}+\frac{d(N^{\prime }m\mathbf{v}^{\prime })}{dt}=N%
\mathbf{f}+N^{\prime }\mathbf{f}^{\prime }=\mathbf{0}  \label{p7}
\end{equation}
where $\mathbf{f}$ represents the force exerting on each constituent of Bob,
likewise $\mathbf{f}^{\prime }$ on each constituent of Alice. Eq.(\ref{p7})
has expressed the law of action and reaction. So we get

\textit{Consequence 3: The newton's third law of motion can also be derived from the
axiom 1 for composite system.}

Even if composite particle is not composed of identical constituent
particles, the Consequence 3 is also valid. Because it is always possible to
divide composite particle into many identical units, each unit is so
small enough that the each unit has the same mass, it is not necessary for
the unit to be a real elementary particle.

\section{An approach to universal gravitational force}

We continue to consider the composite system composed of two particles Bob
and Alice, If they do not affected by external force, their motions obey the
Newton's second law of motion as

\begin{equation}
Bob:N\frac{d(m\mathbf{v)}}{dt}=N\mathbf{f}\qquad \quad N\frac{d(\frac 12mv^2)%
}{dt}=N\mathbf{f}\cdot \mathbf{v}  \label{f6}
\end{equation}

\begin{equation}
Alice:N^{\prime }\frac{d(m\mathbf{v}^{\prime })}{dt}=N^{\prime }\mathbf{f}%
^{\prime }\qquad \quad N^{\prime }\frac{d(\frac 12mv^{\prime 2})}{dt}%
=N^{\prime }\mathbf{f}^{\prime }\cdot \mathbf{v}^{\prime }  \label{f7}
\end{equation}
Let vector $\mathbf{r}$ denote the position of Bob with respect to Alice,
according to the Newton's third law of motion, $N\mathbf{f}+N^{\prime }%
\mathbf{f}^{\prime }=\mathbf{0}$, we obtain that $\mathbf{f}$ (or $\mathbf{f}%
^{\prime }$) parallel or anti-parallel $\mathbf{r}$.

Vector-multiplying Eq.(\ref{f6}) by $\mathbf{r}$, because $\mathbf{f}$
parallels $\mathbf{r}$ , we have

\begin{equation}
\mathbf{r\times [}N\frac{d(m\mathbf{v)}}{dt}]=Nm\frac{d(\mathbf{r}\times 
\mathbf{v})}{dt}=N\mathbf{r}\times \mathbf{f=0}  \label{f14}
\end{equation}
It means

\begin{equation}
\mathbf{r}\times \mathbf{v}=\mathbf{h}=const.  \label{f15}
\end{equation}
where $\mathbf{h}$ is an integral constant. Likewise for Alice. Using $%
\mathbf{f}\parallel \mathbf{r}$, we can expand $\mathbf{f}$ in a Taylor
series in $1/r$, this gives

\begin{equation}
\mathbf{f}=\frac{\mathbf{r}}r(b_0+b_1\frac 1r+b_2\frac 1{r^2}+b_3\frac
1{r^3}+...)  \label{f16}
\end{equation}
Substituting into Eq.(\ref{f6}), we obtain

\begin{eqnarray}
\frac 12mv^2 &=&\int (\mathbf{f}\cdot \mathbf{v})dt=\int |\mathbf{f|}dr 
\nonumber \\
&=&\varepsilon +b_0r+b_1\ln r-b_2\frac 1r-b_3\frac 1{2r^2}-...  \label{f17}
\end{eqnarray}
where $\varepsilon $ is an integral constant. Now consider Eq.(\ref{f15}),
it means that Bob moves around Alice (no matter by attractive or repulsive
interaction), at the perihelion point we find

\begin{eqnarray}
h^2 &=&|\mathbf{r}\times \mathbf{v}|^2=r^2v_\varphi ^2|_{perihelion} 
\nonumber \\
&=&\frac{2r^2}m(\varepsilon +b_0r+b_1\ln r-b_2\frac 1r-b_3\frac
1{2r^2}-...)|_{perihelion}  \nonumber\\
\label{f18}
\end{eqnarray}

Since $b_i$ are the coefficients that are independent from distance $r$,
integral constant $h$ and integral constant $\varepsilon $, they take the
same values for various cases which have various man-controlled parameters $h
$ and $\varepsilon $. Now we consider two extreme cases.

\textit{First case: Bob is at rest forever.}

According to the Newton's first law of motion, the interaction between them
must completely vanishes. Since Bob' speed $v$ should not depend on the
distance, according to Eq.(\ref{f17}), a reasonable solution may be $%
r=\infty $, $b_0=0$, $b_1=0$ and $\varepsilon =0$. To note that the values
of $b_0$ and $b_1$ do not depend on this extreme case. 

\textit{Second case: with }$h\rightarrow 0$, \textit{Bob passes the
perihelion point about Alice with a distance }$r\rightarrow 0$\textit{.}

According to Eq.(\ref{f18}), a reasonable solution may be that the all
coefficients $b_i$ are zero but except $b_2$ and $b_3\rightarrow 0$.
Because $b_3$ does not depend on $h$, we find $b_3=0$.

According to the two extreme cases, we obtain a general expression given by

\begin{eqnarray}
\mathbf{f} &=&b\frac{\mathbf{r}}{r^3}  \label{f19} \\
\frac 12mv^2 &=&\varepsilon +b\frac 1r  \label{f20}
\end{eqnarray}
where the subscript of $b_2$ has been dropped. Likewise for Alice, we have

\begin{eqnarray}
\mathbf{f}^{\prime } &=&a\frac{\mathbf{r}}{r^3}  \label{f21} \\
\frac 12mv^{\prime 2} &=&\varepsilon ^{\prime }+a\frac 1r  \label{f22}
\end{eqnarray}
where $a$ and $\varepsilon ^{\prime }$ are coefficients. To note that the
above four expressions do not rely on whether the situation is extreme case.

Substituting Eq.(\ref{f19}) and Eq.(\ref{f21}) into $N\mathbf{f}+N^{\prime }%
\mathbf{f}^{\prime }=\mathbf{0}$, we get

\begin{equation}
N\mathbf{f}+N^{\prime }\mathbf{f}^{\prime }=N\frac{b\mathbf{r}}{r^3}%
+N^{\prime }\frac{a\mathbf{r}}{r^3}=0  \label{f24}
\end{equation}
This equation leads to

\begin{equation}
\frac b{N^{\prime }}=-\frac aN=K  \label{f25}
\end{equation}
where $K$ is a coefficient. Then Eq.(\ref{f6}) and Eq.(\ref{f7}) may be
rewritten as

\begin{equation}
Bob:\quad N\frac{d(m\mathbf{v)}}{dt}=K\frac{NN^{\prime }\mathbf{r}}{r^3}%
\qquad \qquad  \label{f27}
\end{equation}

\begin{equation}
Alice:\quad N^{\prime }\frac{d(m\mathbf{v}^{\prime })}{dt}=-K\frac{%
NN^{\prime }\mathbf{r}}{r^3}\qquad \qquad   \label{f28}
\end{equation}
As the most simple case, the constituents have identical mass, so the number 
$N$ of constituents in Bob and $N^{\prime }$ in Alice can directly represent
the mass of Bob and the mass of Alice, respectively. If $K$ takes a negative
constant, then, the above equations show that Bob is attracted by Alice with
the Newton's universal gravitational force. Thus we get

\textit{Consequence 4: the Newton's universal gravitational force can be derived
from the axiom 1. But the nature of being attractive or repulsive depends on
experiments.}

Even if composite particles are not composed of identical constituent
particles, the Consequence 4 is also valid, the reason is the same as in the
preceding section.

\subsection{An approach to Coulomb's force}

\label{Coulf}In this section we give an explanation of Coulomb's force by
using the most simple model: all particles are composed of identical
constituents.

From the above section, now we can manifestly interpret the quantity $%
\mathbf{f}$ as the force exerting on a constituent of Bob. It is a natural
idea to think of that constituents have two kinds of charges: positive and
negative. If Bob and Alice are separated by a far distance, and $\mathbf{f}$
is the force acting on a positive constituent in Bob, then $-\mathbf{f}$ is
the force acting on a negative constituent in Bob. Regardless of the
internal forces in Bob,  it follows from Eq.(\ref{f6}) that the motion of
the ith constituent in Bob is governed by

\begin{equation}
\frac{d(m\mathbf{v}^{(i)})}{dt}=\mathbf{f}^{(i)}\quad \ \quad \frac{d(\frac
12mv^{(i)2})}{dt}=\mathbf{f}^{(i)}\cdot \mathbf{v}^{(i)}  \label{c1}
\end{equation}
where $\mathbf{v}^{(i)}$ and $\mathbf{f}^{(i)}$ denote the velocity and the
force acting on the ith constituent, respectively, $m$ denote the mass of a
constituent. Summing over all constituents in Bob, we get

\begin{eqnarray}
\sum\limits_{i=1}^N\frac{d(m\mathbf{v}^{(i)})}{dt} &=&\frac
d{dt}[m\sum\limits_{i=1}^N\mathbf{v}^{(i)}]=\frac{d(Nm\mathbf{v}_c)}{dt}
\label{c2} \\
\sum\limits_{i=1}^m\mathbf{f}^{(i)} &=&q\mathbf{f}_c  \label{c3}
\end{eqnarray}
where $\mathbf{v}_c$ is the central velocity of Bob, $q$ denotes the net
charge number of Bob, $\mathbf{f}_c$ denotes the force acting on the
constituent which locates at the geometric center of Bob ( this central
constituent may be virtual one because it features the average action), we
obtain

\begin{equation}
\frac{d(Nm\mathbf{v}_c)}{dt}=q\mathbf{f}_c\quad  \label{c4}
\end{equation}

Like that in the above section, the Newton's laws of motion must be valid
for the whole composite system of Bob and Alice, in other words, when they
are separated from a infinite distance they are isolated, whereas they go to
nearest point they should not touch each other, these requirements lead to

\begin{eqnarray}
Bob &:&\mathbf{f}_c=b\frac{\mathbf{r}}{r^3}\qquad \frac 12mv^2=\varepsilon
+b\frac 1r  \label{c5} \\
Alice &:&\mathbf{f}_c^{\prime }=a\frac{\mathbf{r}}{r^3}\qquad \frac
12mv^{\prime 2}=\varepsilon ^{\prime }+a\frac 1r  \label{c6}
\end{eqnarray}
where $\varepsilon ,\varepsilon ^{\prime },$ $b$ and $a$ are coefficients.
According to the Newton's third law of motion, we have $q\mathbf{f}%
+q^{\prime }\mathbf{f}^{\prime }=\mathbf{0}$, where $q^{\prime }$ denotes
the net charge number of Alice. We get

\begin{equation}
q\mathbf{f}_c+q^{\prime }\mathbf{f}_c^{\prime }=q\frac{b\mathbf{r}}{r^3}%
+q^{\prime }\frac{a\mathbf{r}}{r^3}=0  \label{c8}
\end{equation}
This equation leads to

\begin{equation}
\frac b{q^{\prime }}=-\frac aq=k  \label{c9}
\end{equation}
where $k$ is a constant. Then the motions of Bob and Alice are governed by

\begin{equation}
Bob:\quad N\frac{d(m\mathbf{v)}}{dt}=k\frac{qq^{\prime }\mathbf{r}}{r^3}%
\qquad \qquad  \label{c10}
\end{equation}

\begin{equation}
Alice:\quad N^{\prime }\frac{d(m\mathbf{v}^{\prime })}{dt}=-k\frac{%
qq^{\prime }\mathbf{r}}{r^3}\qquad \qquad  \label{c11}
\end{equation}
The Eq.(\ref{c10}) and Eq.(\ref{c11}) are known as the Coulomb's forces.

\textit{Consequence 5: Coulomb's force can be derived from the axiom 1.}

Even if composite particles are not composed of identical constituent
particles, the Consequence 5 is also valid. Because it is always possible to
divide composite particle into many identical units, each unit is so
small enough that each unit having the same mass is assigned a charge unit
by keeping the net charges unchanged, it is not necessary for the unit to be
a real elementary particle.

\section{An approach to relativistic mechanics}

Again consider a particle of rest mass $m$ in our frame of reference $%
S(x_1,x_2,x_3,t)$, the particle moves a distance $\Delta l$ during
infinitesimal time interval $\Delta t$ with speed $v$, according to
Pythagoras theorem, we have

\begin{eqnarray}
\Delta l_{}^2 &=&\Delta x_1^2+\Delta x_2^2+\Delta x_3^2 =v^2\Delta t^2 
\nonumber \\
&=&v^2\Delta t^2-c^2\Delta t^2+c^2\Delta t^2  \nonumber \\
&=&c^2\Delta t^2(1-v^2/c^2)+c^2\Delta t^2  \label{p17}
\end{eqnarray}
where $c$ is the speed of light. Defining the modified velocity

\begin{eqnarray}
u_1 &=&\frac{v_1}{\sqrt{1-v_{}^2/c^2}}\qquad u_2=\frac{v_2}{\sqrt{%
1-v_{}^2/c^2}}  \label{p18} \\
u_3 &=&\frac{v_3}{\sqrt{1-v_{}^2/c^2}}\qquad u_4=\frac{ic}{\sqrt{1-v_{}^2/c^2%
}}  \label{p19}
\end{eqnarray}
where $v_{}^2=v_1^2+v_2^2+v_3^2$, from Eq.(\ref{p17}), we obtain

\begin{equation}
u_1^2+u_2^2+u_3^2+u_4^2=-c^2\qquad u_\mu u_\mu =-c^2  \label{p20}
\end{equation}
where the repeated Greek indices take summation over values 1, 2, 3 and 4.
The 4-vector velocity $u=\{u_\mu \}$ is known as the relativistic velocity%
\cite{Harris}. It is convenient to define proper time interval $d\tau =dt%
\sqrt{1-v_{}^2/c^2}$, thus the relativistic velocity is given by

\begin{equation}
u_\mu =dx_\mu /d\tau  \label{p21}
\end{equation}
where $x_4=ict$, Eq.(\ref{p20}) is the magnitude formula of relativistic
4-vector velocity of particle in Minkowski's space in its square form.

(1) The motion of the particle satisfies Eq.(\ref{p20}), we have

\begin{equation}
mu_\mu mu_\mu =-m^2c^2  \label{p22}
\end{equation}
By defining the momentum $P_\mu =mu_\mu $, $\mathbf{P}=(P_1,P_2,P_3)$, we
have

\begin{equation}
P_\mu P_\mu =\mathbf{P}\cdot \mathbf{P}+P_4^2=-m^2c^2  \label{p23}
\end{equation}
Differentiating Eq.(\ref{p23}) with respect to proper time interval $d\tau $%
, we get

\begin{equation}
\mathbf{P}\cdot (\frac{d\mathbf{P}}{d\tau })+P_4\frac{dP_4}{d\tau }=0
\label{p24}
\end{equation}
By using $P_4=mu_4=micdt/d\tau =$ $mic/\sqrt{1-v_{}^2/c^2}$, and introducing
a 3-dimensional vector symbol $\mathbf{F}$, the above equation becomes

\begin{eqnarray}
\mathbf{F} &=&\frac{d\mathbf{P}}{d\tau }  \label{p25} \\
-ic\frac{d(P_4)}{d\tau } &=&\mathbf{F}\cdot \mathbf{v}  \label{p26}
\end{eqnarray}
Eq.(\ref{p26}) can be understood as: the right side represents the power of
the force $\mathbf{F}$ which exerts on the particle with the usual
3-dimensional velocity $\mathbf{v}$; the left side represents the change
rate of the energy $E$ of the particle, i.e.

\begin{equation}
E=-icP_4=mc^2\frac{dt}{d\tau }=\frac{mc^2}{\sqrt{1-v_{}^2/c^2}}  \label{p27}
\end{equation}
Obviously, Eq.(\ref{p27}) is just the relativistic energy of the particle in
relativistic mechanics, Eq.(\ref{p25}) and Eq.(\ref{p26}) are just the
relativistic Newton's second law\cite{Harris1}. Further more, the mass $m$
is called the rest mass of the particle while

\begin{equation}
m_r(v)=\frac m{\sqrt{1-v_{}^2/c^2}}  \label{p28}
\end{equation}
is called the relativistic mass of the particle. This leads to $E=m_rc^2$.
By rewriting, Eq.(\ref{p23}) is given by

\begin{equation}
E^2=|\mathbf{P|}^2c^2+m^2c^4  \label{p29}
\end{equation}
Eq.(\ref{p29}) is just the Einstein's relationship of energy and momentum.

(2) If we define symbols

\begin{equation}
\cosh \alpha _\mu \equiv \frac{u_\mu }{ic}=\frac{dx_\mu }{icd\tau }
\label{p30}
\end{equation}
Eq.(\ref{p20}) can be rewritten as

\begin{eqnarray}
\cosh ^2\alpha _1+\cosh ^2\alpha _2+\cosh ^2\alpha _3+\cosh ^2\alpha _4 &=&1
\nonumber \\
or\qquad \cosh \alpha _\mu \cosh \alpha _\mu &=&1  \label{p31}
\end{eqnarray}
The above equation indicates that there is a definite like-Euclidean
trigonometry in the Minkowski's space, despite all that the fourth axis is
imaginary, the quantity $\alpha _\mu $ may be understood as the angle
between $dx_\mu $ and $icd\tau $.

As the most simple case, only consider the motion taking place in the plane $%
(x_1,x_4)$, the length $icd\tau $ makes angles $\alpha _1$ and $\alpha _4$
with respect to the lengths $dx_1$ and $dx_4$ respectively, we have

\begin{equation}
\cosh ^2\alpha _1+\cosh ^2\alpha _4=1  \label{p32}
\end{equation}
The above equation indicates that the two angles $\alpha _1$ and $\alpha _4$
have formed a ''right-angled triangle'' in the plane $(x_1,x_4)$. By
defining symbol

\begin{equation}
\sinh \alpha _4=\cosh \alpha _1  \label{p33}
\end{equation}
we get

\begin{equation}
\sinh ^2\alpha _4+\cosh ^2\alpha _4=1  \label{p34}
\end{equation}
From Eq.(\ref{p30}), we know

\begin{eqnarray}
\cosh \alpha _4 &=&\frac{u_4}{ic}=\frac 1{\sqrt{1-v_{}^2/c^2}}  \label{p35}
\\
\sinh \alpha _4 &=&\cosh \alpha _1=\frac{u_1}{ic}=\frac{v_{}}{ic\sqrt{%
1-v_{}^2/c^2}}  \label{p36}
\end{eqnarray}
where $v=v_1=dx_1/dt$ for the simple case.

(3)In the Minkowski's space, if the coordinate system $S(x_1,x_2,x_3,x_4)$
"rotates" through an angle $\alpha $ in the plane $(x_1,x_4)$, and becomes
another new coordinate system $S^{\prime }(x_1^{\prime },x_2^{\prime
},x_3^{\prime },x_4^{\prime })$. According to Eq.(\ref{p34}), the
transformation of the two systems $S$ and $S^{\prime }$ will be given by

\begin{eqnarray}
x_1^{\prime } &=&x_1\cosh \alpha -x_4\sinh \alpha  \label{p37} \\
x_2^{\prime } &=&x_2  \nonumber \\
x_3^{\prime } &=&x_3  \nonumber \\
x_4^{\prime } &=&x_1\sinh \alpha +x_4\cosh \alpha  \label{p38}
\end{eqnarray}
By comparison, we know that this $\alpha $ is just that $\alpha _4$ of Eq.(%
\ref{p35}). Substituting Eq.(\ref{p35}) and Eq.(\ref{p36}) into the above
equations, and using $x_4=ict$ and $x_4^{\prime }=ict^{\prime }$, we get

\begin{eqnarray}
x_1^{\prime } &=&\frac{x_1}{\sqrt{1-v_{}^2/c^2}}-\frac{vt}{\sqrt{1-v_{}^2/c^2%
}}  \label{p39} \\
x_2^{\prime } &=&x_2  \nonumber \\
x_3^{\prime } &=&x_3  \nonumber \\
t^{\prime } &=&-\frac{x_1v/c^2}{\sqrt{1-v_{}^2/c^2}}+\frac t{\sqrt{%
1-v_{}^2/c^2}}  \label{p40}
\end{eqnarray}

(4)If Bob is at rest at the origin of the system $S^{\prime }$, i.e. $%
x_1^{\prime }=0$, from Eq.(\ref{p39}), we know that Bob will be moving at
speed $v=dx_1/dt$ in the system $S$. In other words, the system $S^{\prime }$
is fixed at Bob, while Bob is moving at the speed $v$ along the axis $x_1$
of the system $S$. Coupling with this explanation, Eq.(\ref{p39}) and Eq.(%
\ref{p40}) are just the Lorentz transformation.

(5)The constancy of the speed of light, length contraction and time dilation
can evidently be derived from the Lorentz transformation equations.

Therefore, we get

\textit{Consequence 6: The formula of relativistic dynamics can be derived from
the axiom 1.}

\section{Two kinds of 4-vector forces}

(1) In Minkowski's space, according to Eq.(\ref{p25}) and Eq.(\ref{p26}), we
can define a 4-vector force as

\begin{eqnarray}
f &\equiv &(\mathbf{F},f_4)  \label{d1} \\
f_4 &=&-\frac{\mathbf{F}\cdot \mathbf{u}}{u_4}  \label{d2}
\end{eqnarray}
then Eq.(\ref{p25}) and Eq.(\ref{p26}) may be rewritten as

\begin{equation}
f=\frac{d(mu)}{d\tau }\qquad or\qquad f_\mu =\frac{d(mu_\mu )}{d\tau }
\label{d3}
\end{equation}
Obviously, the 4-vector force $f$ and 4-vector velocity $u$ satisfies
orthogonal relation\cite{Rindler}

\begin{eqnarray}
u\cdot f &=&u_\mu f_\mu =u_\mu \frac{d(mu_\mu )}{d\tau }  \nonumber \\
&=&m\frac{d(u_\mu u_\mu )}{d\tau }=m\frac{d(-c^2)}{d\tau }=0  \label{d4}
\end{eqnarray}
This result is easy to be understood when we note that the magnitude of the
4-vector velocity keeps constant, i.e. $|u|=\sqrt{u_\mu u_\mu }=ic$, any
4-vector force can never change the magnitude of the 4-vector velocity but
can change its direction in the Minkowski's space.

The orthogonal relation of 4-vector force and 4-vector velocity is a very
important feature for basic interactions\cite{Cui}.

(2) In analogy with the relativistic mechanics, Newton's mechanics may also
have its own 4-vector force.

According to Eq.(\ref{p3}) and Eq.(\ref{p4}), we define 4-vector quantities
in Newton's mechanics as

\begin{eqnarray}
\widetilde{v} &\equiv &(\mathbf{v},iv)=(v_1,v_2,v_3,iv)  \label{d5} \\
\widetilde{f} &\equiv &(\mathbf{f},\widetilde{f}_4)=(\mathbf{f},-\frac{%
\mathbf{f}\cdot \mathbf{v}}{\widetilde{v}_4})=(\mathbf{f},-\frac{\mathbf{f}%
\cdot \mathbf{v}}{iv})  \label{d6} \\
\widetilde{f} &=&\frac{d(m\widetilde{v})}{dt};\qquad \widetilde{f}_\mu =%
\frac{d(m\widetilde{v}_\mu )}{dt}  \label{d7} \\
\widetilde{v}\cdot \widetilde{v} &=&\widetilde{v}_\mu ~\widetilde{v}_\mu
=0;\quad \widetilde{v}\cdot \widetilde{f}=\widetilde{v}_\mu ~\widetilde{f}%
_\mu =0  \label{d8}
\end{eqnarray}
where the magnitude of the 4-vector velocity $\widetilde{v}$ keeps the
constant zero, so that the 4-vector force $\widetilde{f}$ and the 4-vector
velocity $\widetilde{u}$ satisfies the orthogonal relation of Eq.(\ref{d8}),
while Eq.(\ref{d7}) represents the Newton's second law in the 4-vector form.

(3) Both Newton's mechanics and relativistic mechanics can be derived from
the same Pythagoras theorem in rigorous manner, but modern physics have
clearly show that it prefers to the relativistic mechanics, it was said that
the newton' mechanics is an excellent approximation when the motion is
slow with respect to the speed of light.

If we use the speed of sound to replace the speed of light in
Eq.(\ref{p17}), we would obtain a dynamics resemble to the relativistic
mechanics, in which the speed of sound takes place the role of the speed of
light, how do physics accept it?

Generally speaking, we have absolute confidence on Pythagoras theorem, we
have no doubts on the preceding results which have realize the
geometrization of physics, we must equally treat with the Newton's mechanics
and the relativistic mechanics, rather than that one is another's
approximation. The author is not intent to shake the established modern
physical contents but to pursue the faith coming from the Pythagoras
theorem, the Pythagoras theorem is also an well established law which is at
least 2500 years old.

Why the nature seems to prefer to the relativistic mechanics ? We here gives a
trial explanation. In one hand, to note that all measurements about distance
and time must use facilities whose principles are directly or indirectly
based on the light, therefore, all phenomena in our optical eyes or
instruments have had to involve the speed of light, so that the action we
perceive is the force $\mathbf{F}$ of Eq.(\ref{p25}) rather than the force $%
\mathbf{f}$ of Eq.(\ref{p3}), hence the nature we perceive is the would
obeying the relativistic mechanics. In another hand, if our measurements are
accomplished by virtue of infinite communicating speed (if it exists ), 
the nature we perceive will be the world obeying the Newton's
mechanics. This explanation is completely compatible with the standard
contents of modern physics, because we knew in textbooks\cite{Mulligan}\cite
{Olenick} that relativistic mechanics reduces to Newton's mechanics when $c$
becomes infinite, proofing that our effort is not intent to offend the
established physics but to provide a new insight into the subject.

\section{ Discussion}

(1) The six theorem derived from Pythagoras theorem in the preceding
sections form the fundamentals of mechanics, from them other useful theorems,
such as the Lagrange's Equation and Hamilton's principle, etc., can also be
derived out. With this results, we recognize that the Pythagoras theorem is
the origin of the whole mechanics. It becomes possible to establish an
axiomatical system for the subject, both for teaching and research,
especially, the experiments addressing to to test the mechanics may be
canceled from our classroom because an axiomatical system couldn't tolerate
the existence of many initiations in the subject.

(2) What is the inertial frame of reference? Our answer is that the frame in
which Pythagoras theorem is valid is just the inertial frame of reference.
This answer has a little different from that in traditional textbooks.

In an accelerated frame of reference, Pythagoras theorem needs a
modification, this situation open a way leading to the subject of general
theory of relativity.

(3) Eq.(\ref{f16}) expresses the interaction expanded in a Taylor series in $%
1/r$. letting it contain a Taylor series in $r$ is not necessary, because
the $r$ series would be canceled when we need the interaction vanishes for $%
r\rightarrow \infty $.

(4) Pythagoras theorem has a long history, after Greek
philosopher Pythagoras ( 500 B.C. ). Other ancient civilizations also have the same
theorem, such as India ( 500 B.C.---200 B.C.)\cite{Still}, China ( 1100
B.C.---200 B.C. )\cite{Li}. In China, Pythagoras theorem is called the GouGu theorem, even
today. Both western and eastern share the common wisdom of their ancient
civilizations.

\section{Conclusion}

In this paper, we regard Pythagoras theorem as an axiom, from this axiom we
derived out six important theorems of physics, they are Newton's three laws
of motion, universal gravitational force, Coulomb's force, and the formula
of relativistic dynamics. The results indicate that at least the mechanics
is possible to be geomertrized, so that it is possible to string together
these theorems with Pythagoras theorem. Knowing the internal relationships
between them, which have never been clearly revealed by other author, will
benefit our physics teaching.


\begin{thebibliography}{9}
\bibitem{Harris}  E. G. Harris, Introduction to Modern Theoretical Physics,
Vol.1, John Wiley \& Sons, USA, (1975)263, Eq.(10-39).

\bibitem{Harris1}  See, ref.\cite{Harris}, p.264, Eq.(10-44).

\bibitem{Rindler}  W. Rindler, Essential Relativity, (Springer-Verlag, 2nd
Ed., 1977), p.89.

\bibitem{Cui} H. Y. Cui, Eprint-arXiv: physics/0102073, Feb., 2001.

\bibitem{Mulligan}  J. F. Mulligan, Introductory College Physics,
(McMraw-Hill, USA, 1985), p.735.

\bibitem{Olenick}  R. P. Olenick, T. M. Apostol and D. L. Goodstein, Beyond
The Mechanical Universe, (California Institute of Technology, USA, 1995),
p.389.

\bibitem{Still}  J. Stillwell, Mathematics and its History, (Springer-Verlag,
NY, 1989), p.3.

\bibitem{Li}  W. L. Li, A Source Book in Mathematics, (Science, Beijing,
1998), p.25.
\end{thebibliography}
\end{document}